\documentstyle[aps,multicol,epsf]{revtex}

\begin{document}

\title{
Clustering of correlated networks
}

\author{
S.N. Dorogovtsev$^{1, 2, \ast}$  
}

\address{
%$^{1}$ A.F. Ioffe Physico-Technical Institute, 194021 St. Petersburg, Russia\\
$^{1}$ Departamento de F\'\i sica and Centro de F\'\i sica do Porto, Faculdade 
de Ci\^encias, 
Universidade do Porto,\\
Rua do Campo Alegre 687, 4169-007 Porto, Portugal
\\
$^{2}$ A.F. Ioffe Physico-Technical Institute, 194021 St. Petersburg, 
Russia 
}

\maketitle

\begin{abstract} 
We obtain the clustering coefficient, the degree-dependent local clustering, and the mean clustering of networks with arbitrary correlations between the degrees of the nearest-neighbor vertices. The resulting formulas 
allow one to determine the nature of the clustering of a network. 
\end{abstract}


\begin{multicols}{2}

\narrowtext

%%%%%%%%%%%%%%%%%%%%%%%%%%%%%%%%%%%%%%%%%%%%%%%%%%%%%%%%%%%%%%%%%%%%%%% 

In principle, loops, and, in particular, loops of length three which lead to the clustering of networks, are a specific kind of correlations. Usually, real-world networks are strongly clustered structures, and many efforts were made to invent special mechanisms producing strong clustering even in small nets \cite{w99,s01}. The number of the proposed mechanisms is rapidly growing, but the recent development of the field 
\cite{ab02,dm02,bookdm03,n03} shows that in very many real networks the high clustering is only a finite-size effect. So, in this case, no additional mechanism of strong clustering is needed. 
The problem is to reliably conclude whether or not the clustering of a real network is a finite-size effect 
which can be explained by using basic random graph constructions 
\cite{n02b}. Evidently, comparison with results obtained in the framework of specific models with many adjusting parameters cannot lead to any convincing conclusion.  
%%The reliable conclusion must be based on      

Another basic though particular kind of correlations in networks are correlations between the numbers of connections (degrees) of the nearest neighbor vertices \cite{kr01,pvv01,vpv02,ms02,msz02,msa02,n02,bl02,ccrm02,pn03,vbmpv03,tmms03,d03}. Networks with these specific correlations are being extensively studied these days, and the term ``correlated networks'' often implies just this type of correlations. These pair correlations were measured in a number of real networks \cite{pvv01,vpv02,ms02,msz02,msa02,n02,tmms03}, so the joint distribution of the degrees of the nearest neighbor vertices, $P(k,k')$ is considered as one of metrics of a network. Note that as a rule, these correlations do not vanish in the large network limit. 

The classical random graphs \cite{er59,sr51} with their Poisson degree distribution provide a non-adequate image of a real complex network and a very weak 
clustering, $C = \overline{k}/N$. Here $\overline{k}$ is the mean degree of a graph and $N$ is its size (the total number of vertices). 
Random graphs with given degree-distribution $P(k)$ (the configuration model of mathematical graph theory \cite{bbk72}) are much closer to real complex networks. It is the values of the clustering coefficient of this model $C \propto N^{-1}$ \cite{n02b,emb02} 
that were compared with empirical data for real-world networks.  
  
The configuration model and its variations provide (uncorrelated) random graphs which are maximally random (i.e., with the maximum entropy) under the constraint that their degree distribution is equal to a given one, $P(k)$. 
These graphs are closer to reality than the classical random graphs, but the absence of correlations is a very restrictive factor. 
If we wish to make a step toward real networks, we have to introduce a network with degree--degree correlations, $P(k,k')$. The simplest formal way to do this is in the spirit of the configuration model. That is, consider random graphs which are maximally random under the constraint that their degree--degree distribution is equal to a given one, $P(k,k')$. 
%%\cite{xxx,xxx}. 
This is the minimal construction of a random graph with these correlations. 
In this construction, as the size of a graph approaches infinity, loops become insignificant, and the clustering vanishes \cite{note1}.  

In the present communication we obtain analytical expressions for the complete list of the clustering characteristics of the random graphs with these important degree--degree correlations, see Eqs. (\ref{10})--(\ref{12}). 
These formulas, after the substitution of a measured distribution $P(k,k')$, allow one to conclude whether or not the clustering of a real-world or generated network is simply a finite-size effect, the same as in a maximally random graph with this degree--degree distribution.  
%%Although the clustering of these correlated graphs behaves as  
%%$\propto N^{-1}$, t 
Furthermore, the resulting clustering characteristics are qualitatively different from those of uncorrelated networks.  

The graphs in this communication are completely described by the joint distribution $P(k,k')$ of the degrees of end vertices of an edge of the graph, $\sum_{k,k'} P(k,k') = 1$, $P(k,k') = P(k',k)$. The degree distribution $P(k)$ is determined by $P(k,k')$: 
%%%
\begin{equation}
P(k) = \frac{\overline{k}}{k}\sum_{k'} P(k,k') 
\, ,      
\label{1}
\end{equation} 
%%% 
where the mean degree $\overline{k} \equiv \langle k \rangle \equiv \sum_k kP(k)$ is  
%%%
\begin{equation}
\overline{k} = \Biggl[ \sum_{k,k'} \frac{P(k,k')}{k} \Biggr]^{-1}
\, .   
\label{2}
\end{equation} 
%%% 
In the following, we assume that the total number of vertices of the graph, $N$, is large and consider only the main contribution to the clustering.

$P(k,k')$ can be obtained by using empirical data as follows. 
If $k \neq k'$, $P(k,k') = P(k',k)$ is one half of the ratio of the number of edges connecting vertices of degrees $k$ and $k'$ to the total number of edges, $L$. $L=\overline{k}N/2$. If $k=k'$, $P(k,k)$ is the ratio of the number of edges connecting vertices of degrees $k$ and $k$ to $L$.   
  
The set of clustering characteristics of networks, considered up to now, includes: 

\noindent 
(i) The degree-dependent local clustering $C(k)$. This is the mean relative number of connections (less than $1$) between two nearest neighbors of a vertex of degree $k$:  
%%%
\begin{equation}
C(k) \equiv
\frac{\langle m_{\mbox{\scriptsize nn}}(k) \rangle }{k(k-1)/2}
\, ,    
\label{3}
\end{equation} 
%%%  
where $\langle m_{\mbox{\scriptsize nn}}(k) \rangle$ is the average number of connections between the nearest neighbors of a vertex of degree $k$. 
%%One may say, this is the probability that two nearest neighbors 
%%of a vertex of degree $k$ are connected. 

\noindent 
(ii) The mean clustering (mean clustering coefficient), which is defined as   
%%%
\begin{equation}
\overline{C} \equiv \sum_k P(k)C(k)
\, .   
\label{4}
\end{equation} 
%%%  

\noindent 
(iii) The clustering coefficient, which is defined as     
%%%
\begin{equation}
C \equiv  
\frac{\sum_k P(k)\langle m_{\mbox{\scriptsize nn}}(k) \rangle}{\sum_k P(k)k(k-1)/2} =
\frac{\sum_k k(k-1)P(k)C(k)}{\langle k^2 \rangle - \overline{k}}
\, .   
\label{5}
\end{equation} 
%%% 
This coincides with the traditional definition: the clustering coefficient is three times the ratio of the total number of loops of length three in a graph to the total number of connected vertex triples. 
In simple terms, this is the ``concentration'' of loops of length three. 

We shall obtain the clustering characteristics $C(k)$, $\overline{C}$, and $C$ of correlated graphs, 
but let us first introduce 
the conditional probability $P(k|k')$ that if one end vertex of an edge is of degree $k'$, then its other end vertex is of degree $k$:  
%%%
\begin{equation}
P(k|k') = \frac{P(k,k')}{\sum_k P(k,k')} = 
\overline{k}\,\frac{P(k,k')}{k'P(k')}
\, .   
\label{8}
\end{equation} 
%%% 
Then the local clustering, that is, the probability that two nearest neighbors of a vertex of degree $k>1$ are connected is 
%%%
\begin{equation} 
C(k) = \sum_{q,q'> 1} 
P(q'|k)P(q|k)\cdot P(q'|q)\frac{(q'-1)}{Nq'P(q')}\cdot (q-1) 
\, .   
\label{9}
\end{equation} 
%%% 
One can easily understand this formula:  

\noindent
(i) The first two factors $P(q'|k)P(q|k)$ on the right-hand side, which should be accounted for before the summation over $q$ and $q'$, are evident: these are the probabilities that the vertices are of degrees $q$ and $q'$. 

\noindent
(ii) In fact, we must  
calculate the probability that the nearest neighbors with degrees $q$ and $q'$ of a vertex of degree $k$ are connected to each other. 
We have two vertices with $q-1$ and $q'-1$ ``free connections'' 
(apart of the connections to the mother vertex). 
Let us select one of the ``free connections'' of the $q$-vertex. 
The probability that this edge will ``choose'' one of the $q'-1$ ``free connections'' of the $q'$-vertex is given by the product between two central dots on the right-hand part of the formula. The factor $F(q'|q)$ is evident: the second end of the edge must be of degree $q'$. So our edge must ``choose'' (``grasp'') one of the $q'-1$ ``free connections'' of the $q'$-vertex among almost 
$Nq'P(q')$ possibilities in the network. (All these possibilities are equiprobable in the construction which is considered here.) This is the total number of ``free connections'' provided by $NP(q')$ vertices of degree $q'$ in the network. This gives $(q'-1)/[Nq'P(q')]$. 

\noindent
(iii) Finally, we must 
multiply this probability by the number $q-1$ of the ``free connections'' of the $q$-vertex. 

The result is Eq. (\ref{9}). Note that we used the fact that $N$ is large and the probability that the edge between the nearest neighbor is present is 
small, so our formulas are asymptotic. 
Substituting Eq. (\ref{8}) into Eq. (\ref{9}) gives the degree-dependent local clustering   
%%%
\begin{eqnarray} 
& & C(k) = \frac{\overline{k}^3}{N k^2 P^2(k)} \times 
\nonumber
\\[5pt]
& & \sum_{q,q'> 1}\!\!\frac{(q'-1)(q-1)P(q',q)P(q',k)P(q,k)}{qq'P(q)P(q')}
\, ,   
\label{10}
\end{eqnarray} 
%%% 
the mean clustering 
%%%
%\begin{eqnarray} 
%& & 
\begin{equation}
\overline{C} = \frac{\overline{k}^3}{N} 
%\times
%\nonumber
%\\[5pt]
%& & 
\sum_{k,q,q'> 1}\!\!\!\!
\frac{(q'-1)(q-1)P(q',q)P(q',k)P(q,k)}{k^2 qq'P(q)P(q')P(k)}
\, ,    
\label{11}
\end{equation}
%\end{eqnarray} 
%%%  
and the clustering coefficient
%%%
\begin{eqnarray} 
& & C = \frac{\overline{k}^3}{N(\langle k^2 \rangle - \overline{k})} \times
\nonumber
\\[5pt]
& & \sum_{k,q,q'> 1}\!\!\!\!
\frac{(k-1)(q'-1)(q-1)P(q',q)P(q',k)P(q,k)}{kqq'P(q)P(q')P(k)}
\,    
\label{12}
\end{eqnarray} 
%%%
of the correlated network with given correlations $P(k,k')$. 
The degree distribution $P(k)$ in these formulas may be expressed in terms of $P(k,k')$ by using the relations (\ref{1}) and (\ref{2}). The results (\ref{10})--(\ref{12}) may be written in a more compact form in terms of conditional probabilities, see Eq. (\ref{8}), but the present form is more convenient for empirical researchers.  

In uncorrelated networks, $P(k,k')=kP(k)k'P(k')/\overline{k}^2$ and the probability that the nearest neighbor of a vertex is of degree $k$ is $kP(k)/\overline{k}$. In this case, Eqs. (\ref{10})--(\ref{12}) reduce to the known 
result \cite{n02b,emb02}  
%%%
\begin{equation}
C(k) = \overline{C} = C = \frac{(\langle k^2 \rangle - \overline{k})^2}{N\overline{k}^3}
\, .   
\label{7}
\end{equation} 
%%% 
The formulas (\ref{10})--(\ref{7}) are asymptotically exact. 

Note that in uncorrelated networks, $C(k)$ is independent of $k$ and so all the three characteristics are equal. 
Cotrastingly, degree--degree correlations lead to 
a degree-dependent local clustering [see Eq.~(\ref{10})]. Previously, this feature was observed in a number of model and real networks \cite{dgm02,sak03,rb03,rsmob03,v03,ffh03}. Here we demonstrate that this dependence and is a direct consequence of degree--degree correlations. The degree-dependent local clustering leads to the difference between $\overline{C}$ and $C$, which were found in many real-world networks \cite{rb03,rsmob03,v03}. 

One should note that the formula (\ref{10}) for the degree dependent local clustering resembles the expression (61) for the local clustering of a correlated network with hidden variables in the recent paper of Mari\'an Bogu\~n\'a and Romualdo Pastor-Satorras, Ref.~\cite{bp03}. However, there is an essential difference between these two results. The result of Ref.~\cite{bp03} is $C(k)$, expressed in terms of the correlations of hidden variables (``fitnesses'') which were used to generate a correlated network. 
It is impossible to find the exact form of these hidden variable correlations from empirical data. 
Contrastingly, Eq.~(\ref{10}) in the present work is obtained for a random network, which is completely described by $P(k,k')$, and expresses $C(k)$ directly in terms of the observable degree--degree distribution $P(k,k')$. It is the latter circumstance that allows one to use Eqs. (\ref{10})--(\ref{12}) for the structural analysis of networks.  

The number of edges connecting vertices of degrees $k$ and $k'$ can be easily measured in any real-world or generated network \cite{ms02,msz02,msa02,tmms03}. Substituting these numbers together with the numbers of vertices of degree $k$ into Eqs. (\ref{10})--(\ref{12}) will provide one with the clustering characteristics of a maximally random graph with the same degree--degree correlations as the real network. These clustering characteristics may be compared with those of 
the real net. If the results are close enough, then the clustering of a net 
is explained by the basic correlated random graph construction and so 
is a simple finite-size effect. Only if the calculated characteristics differs strongly from the measured ones, the clustering has non-trivial nature. 

Note that in sparse networks, measured degree--degree distributions strongly fluctuates due to poor statistics. This factor cannot spoil the results (\ref{10})--(\ref{12}), since even strong fluctuations are summed out.       

One should indicate two restrictions. (i) The formulas (\ref{10})--(\ref{7}) are asymptotic (large $N$, sufficiently ``weak'' clustering). So, one may hope that they are good if $C$ is less than, say, $0.1$, but only qualitative comparison is possible if, e.g., $C \sim 0.3$. (ii) The growth of real-world networks produces a wide spectrum of correlations, and the correlations between the degrees of the nearest-neighbor vertices are only one specific type of the correlations. The construction that is considered in this communication ignores the long-range and multi-vertex correlations. The empirical data on such correlations is absent. 

%%.. for comparison with measured clustering characteristics, much  
%%than Eq.~(\ref{7}) which was used up to now. 
 
In summary, we obtained the clustering characteristics of networks with correlations between degrees of the nearest-neighbor vertices. These correlations are a common feature of real networks. Our formulas allow one to easily conclude whether or not the clustering of a network is determined by the form of its degree--degree distribution and so  
is a simple finite-size effect. 
%%%, and so 
So, Eqs. (\ref{10})--(\ref{12}) can shed light on the nature of the clustering of networks. We hope that these simple expressions will be a useful tool at the analysis of real-world and generated networks. 
\\[-5pt]

This work was  
partially  
supported by the project POCTI/1999/FIS/33141. 
The author thanks A.V.~Goltsev, M. Bogu\~n\'a, and A.N.~Samukhin for 
%%many 
useful discussions.  
%%%%A.N.S. thanks the NATO program OUTREACH for support. 
Special thanks to the Centro de F\'\i sica do Porto and J.F.F.~Mendes.      
\\

\noindent
{\small $^{\ast}$      Electronic address: sdorogov@fc.up.pt} 
%%%%%\\
%%%%%%%%{\small $^{\dagger}$   Electronic address: jfmendes@fc.up.pt} \\
%%%%%{\small $^{\dagger}$ 
%%%%%%%%\ddag }$ 
%%%%%Electronic address: samukhin@fc.up.pt}

\end{multicols} 


\begin{references}
\bibitem{note1} There is a hierarchy of minimal constructions of complex networks: (i) the maximally random network with a given  degree distribution 
$P(k)$; (ii) the maximally random network with a given distribution $P(k,k')$ of the degrees of the nearest-neighbor vertices; (iii) the maximally random network with a given distribution $P(k,k',k'')$ of the degrees of a triple of connected vertices; and so on.  
All these are equilibrium networks with a tree-like local structure. 
In this communication we discuss only random graphs with a given degree-degree distribution $P(k,k')$, since any empirical data on $P(k,k',k'')$ and higher-order multi-vertex correlations is absent.
\end{references}
\end{document}